\documentclass{PoS}

\title{Strange quark contribution to nucleon form factors}

\ShortTitle{Strange quark contribution to nucleon form factors}

\author{\speaker{Ronald~Babich}\hspace{-0.3em},$^{\;a}$ Richard~Brower,$^{ab}$
        Michael~Clark,$^b$ George~Fleming,$^c$ James~Osborn,$^{bd}$
	and Claudio~Rebbi$^{ab}$\\ \\
        \llap{$^a$}Department of Physics, Boston University,\\
	590 Commonwealth Avenue, Boston, MA 02215, USA\\
        \llap{$^b$}Center for Computational Science, Boston University,\\
	3 Cummington Street, Boston, MA 02215, USA\\
        \llap{$^c$}Department of Physics, Yale University,\\
	New Haven, CT 06520, USA\\
	\llap{$^d$}Argonne Leadership Computing Facility,\\
	9700 S. Cass Avenue, Argonne, IL 60439, USA\\ \\
        E-mail: \email{rbabich@bu.edu}, \email{brower@bu.edu},
        \email{mikec@bu.edu}, \email{George.Fleming@yale.edu},
        \email{josborn@bu.edu}, \email{rebbi@bu.edu}}

\abstract{We discuss methods for the calculation of disconnected
diagrams and their application to various form factors of the nucleon.
In particular, we present preliminary results for the strange
contribution to the scalar and axial form factors, calculated with
$N_f=2$ dynamical flavors of Wilson fermions on an anisotropic
lattice.}

\FullConference{The XXV International Symposium on Lattice Field Theory\\
		 July 30 - August 4 2007\\
		 Regensburg, Germany}

\renewcommand{\logo}{
\raisebox{-5mm}{
\parbox{\textwidth}{
\includegraphics{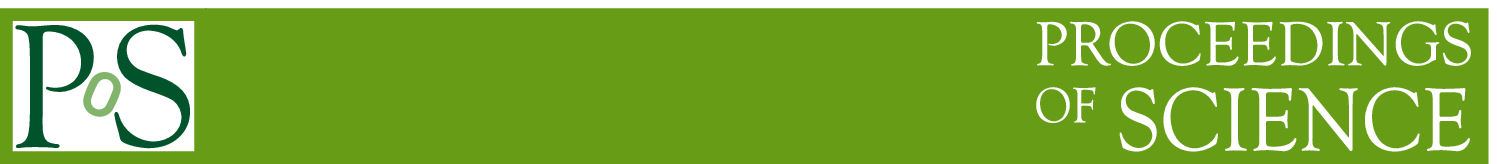}
\begin{flushright}
BUHEP-07-07
\end{flushright} }}}

\begin{document}

\section{Introduction}

In recent years, there has been a great deal of progress in probing
the structure of the nucleon on the lattice.  With few exceptions,
however, such studies have been restricted to the calculation of
isovector quantities or otherwise neglect the contribution of
``disconnected diagrams,'' due to the large cost associated with their
calculation.  The full inclusion of such contributions is necessary,
however, to complete the picture of the nucleon, and with new methods
and the computational resources now available, it appears that the
time may be ripe to do so.

\begin{figure}[b]
\begin{center}
\includegraphics*[width=0.6\textwidth]{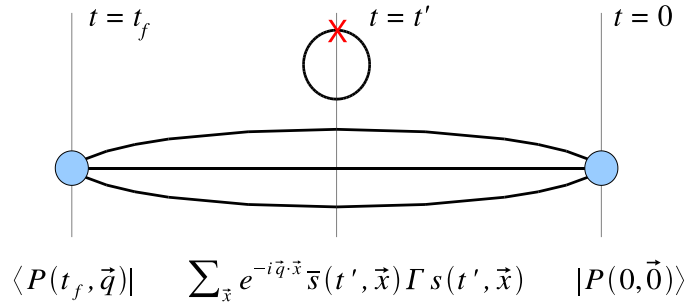}
\caption{\label{fig-disco}Schematic representation of a disconnected
diagram, giving a strange form factor of the proton.  Here $\Gamma$
is the appropriate gamma insertion for the form factor of interest.}
\end{center}
\end{figure}

In this contribution, we focus on a family of observables whose matrix
elements are inherently disconnected, the strange quark contribution
to the elastic form factors of the nucleon.  Such a matrix element is
shown schematically in Fig.~\ref{fig-disco}; by ``disconnected,'' we
mean that the diagram includes an insertion on a quark loop that is
coupled to the baryon correlator only via the gauge field.  This
requires the calculation of a trace of the quark propagator over spin,
color, and spatial indices.  Since an exact calculation would require
a number of inversions proportional to the lattice volume, the trace
is generally estimated stochastically, which introduces a new source
of statistical error whose reduction is discussed in
section~\ref{trace}.  In particular, we propose a novel method for
variance reduction based on the subtraction of the coarse-grid
operator as defined in an adaptive multigrid
scheme.

The strange contribution to the electromagnetic and axial form
factors, $G^s_E(Q^2)$, $G^s_M(Q^2)$, and $G^s_A(Q^2)$, are the subject
of much experimental interest but remain poorly determined to date.
For a review and recent determination, see Ref.~\cite{Pate:2007rf}.  The
scalar form factor, $G^s_S(Q^2)$, is not directly accessible to
experiment, but nevertheless plays an important role in models of
nucleon structure.  Several lattice calculations of these quantities
have been attempted, primarily in the quenched approximation and with
varying degrees of success (see, for example, Ref.~\cite{Fukugita:1994fh,
Fukugita:1994ba,Dong:1995rx,Dong:1995ec,Dong:1997xr,Gusken:1998wy,
Gusken:1999as,Mathur:2000cf,Lewis:2002ix}).  In this
contribution, we discuss the outlook for a calculation of the strange
form factors on large, unquenched, anisotropic lattices.  We also
present preliminary results from an exploratory calculation of the
axial and scalar form factors at $Q^2=0$.

The paper is organized as follows.  In section~\ref{trace}, we discuss
general considerations for calculating the trace and describe the
particular approach employed in our exploratory calculation.  We also
outline a multigrid method for variance reduction.  In
section~\ref{formfact}, we discuss our approach for extracting the
form factors from the corresponding matrix elements.  Finally, in
section~\ref{results} we present preliminary results for
$G^s_A(Q^2=0)$ and $G^s_S(Q^2=0)$.

\section{\label{trace}Trace estimation}

\subsection{Noisy estimators and dilution}

The standard method for estimating the trace relies on calculating the
inverse against a set of noise vectors $\eta$ whose components are
random elements of $U(1)$ or $Z_2$~\cite{Dong:1993pk}, as follows:
\begin{equation}
\mathrm{Tr}(\Gamma D^{-1})\approx \frac{1}{N} \sum_{i=1}^N
\eta_i^\dagger \Gamma D^{-1}\eta_i,\quad \langle\eta_{(x)}^\dagger
\eta_{(y)}\rangle = \delta_{x,y}\,.
\end{equation}
Given a finite ensemble of such noisy sources, this procedure
introduces a new source of statistical error due to the off-diagonal
terms that do not cancel exactly.  Since the fall-off of the quark
propagator is exponential, this error is dominated by the ``near
off-diagonals,'' terms that connect points local in space (we treat
the sum over color and spin exactly).  It follows that one may greatly
reduce this error via {\em dilution}, i.e.~by dividing the stochastic
source into subsets and inverting on these
separately~\cite{Foley:2005ac}.  For example, with simple even/odd
dilution,
\begin{equation}
\mathrm{Tr}(\Gamma D^{-1})\approx \frac{1}{N} \sum_{i=1}^N \eta_i^{(e)\dagger}
\Gamma D^{-1}\eta_i^{(e)}+\frac{1}{N} \sum_{i=1}^N \eta_i^{(o)\dagger}
\Gamma D^{-1}\eta_i^{(o)}\,.
\end{equation}
Here $\eta_i^{(e)}$ and $\eta_i^{(o)}$ are non-zero only on the even
and odd sites, respectively, and $\eta_i^{(e)}+\eta_i^{(o)}=\eta_i$
gives the original noise vector.

In a full calculation, one faces two sources of error: the usual gauge
noise and the error in the trace.  As a baseline in our exploratory
calculation, we largely eliminate the second source of error by
calculating a ``nearly exact'' trace on each of four time-slices.
This is accomplished by employing a large number of sources ($1024
\times 12$ for color/spin) where each source is nonzero on only four
sites on each of the four time-slices.  The sites are chosen such that
the smallest spatial separation between them is $8\sqrt{2}a_s$.  Any
residual contamination, which we observe to be small, is gauge-variant
and averages to zero.  As an aside, we note that this approach is
equivalent to using a single noise vector (or more precisely, a set of
three vectors over space and color that are mutually orthogonal in
color) together with ``extreme dilution.''

\subsection{Multigrid variance reduction}

A number of methods have been proposed to reduce ``far off-diagonal''
contributions to the variance in estimates of the trace
(e.g. Ref.~\cite{Collins:2007mh} in these proceedings).  Such
contributions become increasingly important if we are to consider
light quarks.  A particularly promising approach is to calculate the
trace exactly in the subspace spanned by the lowest eigenmodes of the
Dirac operator~\cite{Neff:2001zr} and estimate the remaining piece
stochastically~\cite{Bali:2005fu}.  In addition to reducing the
variance, this method has the advantage that the set of eigenvectors,
once calculated, may also be used to precondition the inverter and
speed up the large number of inversions needed for the remaining
piece.

Recently, an adaptive geometric multigrid algorithm was shown to
greatly reduce critical slowing down in the two-dimensional U(1)
Schwinger model~\cite{Brannick:2007ue} (see also Ref.~\cite{clarkproc}
in these proceedings).  The method generalizes straightforwardly to
four dimensions, and like eigenvector projection, it is most
advantageous when the set-up cost may be amortized over a large number
of inversions, as is typical for disconnected diagrams.  (The methods
proposed in Refs.~\cite{Luscher:2007se,Stathopoulos:2007zi,morgan07}
have similar advantages.)  Even more promising for this application,
however, is the possibility of greatly reducing the variance in
stochastic estimates of the trace by first subtracting the inverse
calculated on the coarse level.  Following
Ref.~\cite{Brannick:2007ue}, we define a coarse approximation $\tilde
A=P^\dagger A P$ to the positive definite operator $A=D^\dagger D$,
where $P$ is the prolongator that takes vectors from the coarse to the
fine grid and $P^\dagger$ is the corresponding restriction operator.
The inverse $\tilde{A}^{-1}$ can be calculated inexpensively, and
hence the operator \(P\tilde{A}^{-1}P^{\dagger}\) can be used in an
approximation to the full trace, i.e.~$\mathrm{Tr}(\Gamma D^{-1})
\approx \mathrm{Tr}(\Gamma P \tilde{A}^{-1}P^\dagger D^\dagger)$.
Employing the cyclic property of the trace, we have
\(\mathrm{Tr}(\Gamma P \tilde{A}^{-1}P^\dagger D^\dagger) =
\mathrm{Tr}(\tilde{A}^{-1}P^\dagger D^\dagger\Gamma P)\).  This is now
the trace of an operator on the coarse grid, which can be determined
at a much smaller cost, either stochastically or exactly.  At the same
time, it captures the long-range physics and should give a
very good estimate of the full trace.  For an unbiased estimate, we
require the residual contribution on the fine grid, but this
contribution is expected to be small and may be estimated with a smaller
number of (possibly diluted) noise vectors, $N$, without significantly
affecting the variance.  An unbiased estimate of the full trace is
thus given by
\begin{equation}
\mathrm{Tr}(\Gamma D^{-1})\approx
\frac{1}{N}\sum_{i=1}^N\eta_i^\dagger \left[\Gamma
D^{-1}-\Gamma P \tilde{A}^{-1}P^\dagger D^\dagger\right]\eta_i
+\frac{1}{\tilde N}\sum_{i=1}^{\tilde N}\tilde\eta^{\dagger}_i
\left[\tilde{A}^{-1}P^\dagger D^\dagger \Gamma P \right]\tilde\eta_i\,.
\end{equation}
Here $\tilde N$ is the number of noise vectors $\tilde\eta$ on the
coarse grid.  We note, however, that given the large reduction in
degrees of freedom, it will often be practical to calculate the second
term exactly.

The multigrid method of Ref.~\cite{Brannick:2007ue} may be generalized
to work with the Dirac operator $D$ directly, rather than $A=D^\dagger
D$.  In this case the restriction operator is no longer the
conjugate of the prolongation operator, but the basic variance
reduction method goes through as before.  Work is underway to apply
this method as a post-processing step to the results reported below.

\section{\label{formfact}Extracting form factors}

Given an estimate of the trace on each of an ensemble of
configurations, the next step is to correlate these with the proton
correlator to calculate the form factors.  A number of methods exist
for doing so.  Those that have been used to date rely on having an
estimate of the trace on many adjacent time-slices.  In the
most basic approach, the trace is calculated over the entire lattice,
and the form factor is extracted from the time-dependence of the
proton correlator, correlated with this background.  Here we take a
more direct approach by inserting the quark loop on a single
time-slice, labeled by $t'$, at the midpoint of the proton correlator
(see Fig.~\ref{fig-disco}).  The source and sink are moved apart
symmetrically, with $t_f-t'=t'-t_0$, and we look for a plateau at
large separations.  More concretely, for the axial form factor at
$Q^2=0$, we calculate
\begin{eqnarray}
R_A(t,t',t_0) &=& \frac{1}{3}\sum_{i=1}^3\frac{\sum_{\vec x,\vec x'}
[(1+\gamma_4)\gamma_i\gamma_5]^{\alpha\beta}
[\langle P^\beta(\vec x,t) A_i(\vec x',t')\bar{P}^\alpha (\vec 0,t_0)\rangle
-\langle P^\beta(\vec x,t)\bar{P}^\alpha (\vec 0,t_0)\rangle
\langle A_i(\vec x',t')\rangle]}
{\sum_{\vec x}(1+\gamma_4)^{\alpha\beta}\langle P^\beta(\vec x,t)
\bar{P}^\alpha(\vec 0,t_0)\rangle} \nonumber \\
&\rightarrow& G_A^s(Q^2=0)\ \ \mathrm{for}\ \ (t-t')=(t'-t_0)\ \mathrm{large.}
\label{eq-axial}
\end{eqnarray}
Because the expectation value of the axial current vanishes, the
subtraction of the second term is not formally required, but it may
serve to cancel contributions to the error.  For the scalar form
factor, the subtraction of the nonvanishing condensate is necessary,
and we have
\begin{eqnarray}
R_S(t,t',t_0) &=& \frac{\sum_{\vec x,\vec x'} (1+\gamma_4)^{\alpha\beta}
\langle P^\beta(\vec x,t) [\bar\psi \psi(\vec x',t')]
\bar{P}^\alpha (\vec 0,t_0)\rangle} {\sum_{\vec x}(1+\gamma_4)^{\alpha\beta}
\langle P^\beta(\vec x,t)\bar{P}^\alpha(\vec 0,t_0)\rangle}
- \sum_{\vec x'}\langle \bar\psi \psi(\vec x',t')\rangle \nonumber \\
&\rightarrow& G_S^s(Q^2=0)\ \ \mathrm{for}\ \ (t-t')=(t'-t_0)\ \mathrm{large.}
\label{eq-scalar}
\end{eqnarray}
It is important to note that if one has found the trace on multiple
time-slices, this information is not wasted, since one can repeat the
measurement with the system centered at various points in the lattice
and thereby increase the statistics.  For the results presented below,
we have employed four evenly-spaced timeslices.

\section{\label{results}Preliminary results}

At present, we are preparing to calculate disconnected diagrams on
large anisotropic lattices with $N_f=2+1$ dynamical flavors of
clover-improved Wilson fermions.  These lattices are being generated by
members of the Lattice Hadron Physics Collaboration (LHPC) for the
purpose of studying excited-state spectroscopy (see, e.g.,
Ref.~\cite{Lin:2007yf} in these proceedings).  Significant effort is
being invested to construct improved variational sources for the
nucleon on these lattices, and we plan to utilize these to improve
the signal for the disconnected form factors.

In preparation for this calculation, we have performed an exploratory
study on a smaller plain-Wilson test lattice of size $16^3\times 64$
with two dynamical flavors and $M_\pi\approx 360$~MeV.  The lattice is
anisotropic with $a_s=0.118(2)\ \mathrm{fm}\approx 3a_t$, and we have
analyzed 351 independent configurations.  Proton correlators were
calculated with gaussian smearing at source and sink, and we find
that the ground state is isolated near $t=10a$.

\begin{figure}
\begin{minipage}[t]{0.48\textwidth}
\includegraphics*[width=\textwidth]{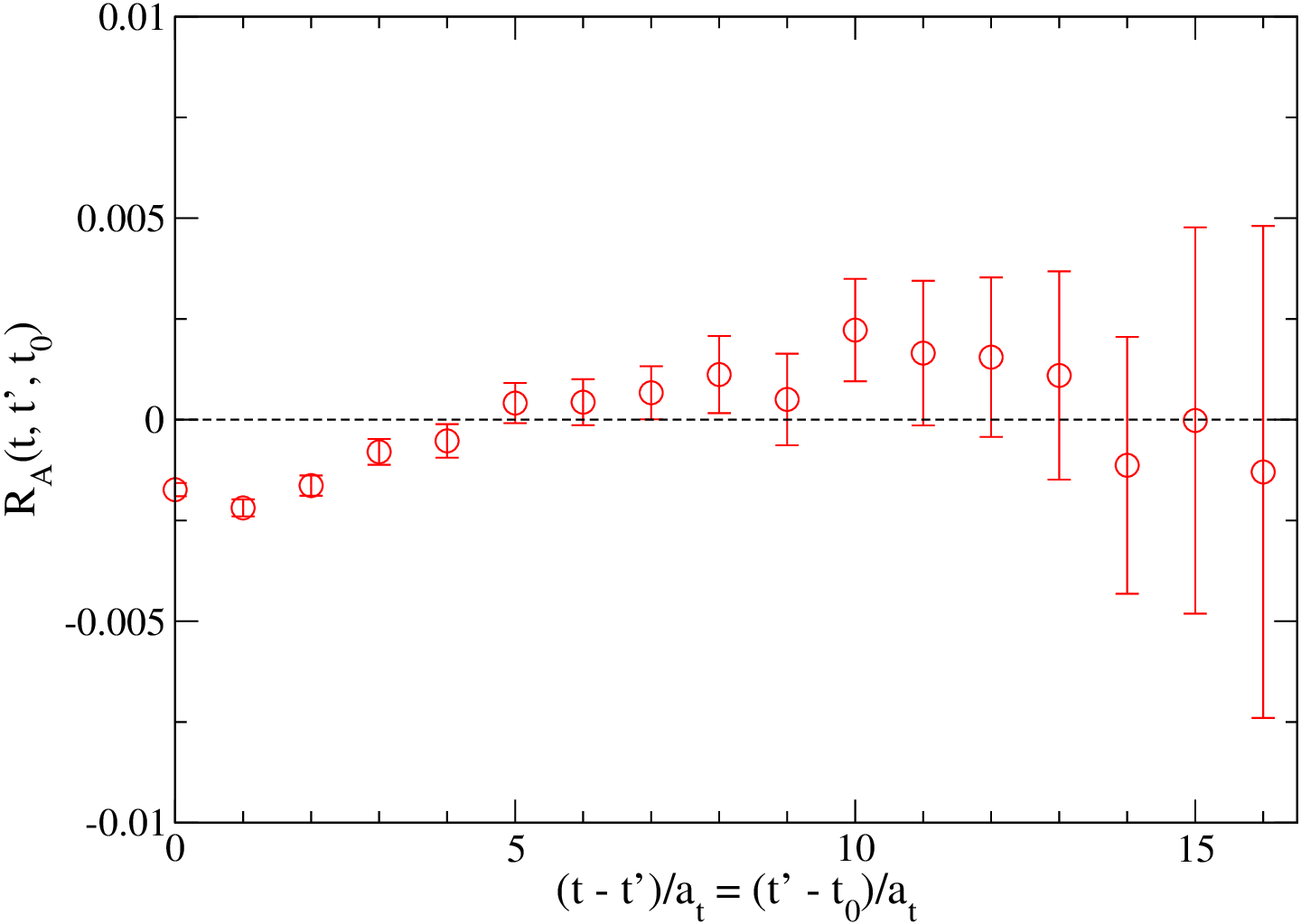}
\caption{\label{fig-axial}Results for axial form factor, with $R_A$ as
defined in the text.}
\end{minipage}
\hfill
\begin{minipage}[t]{0.48\textwidth}
\includegraphics*[width=0.95\textwidth]{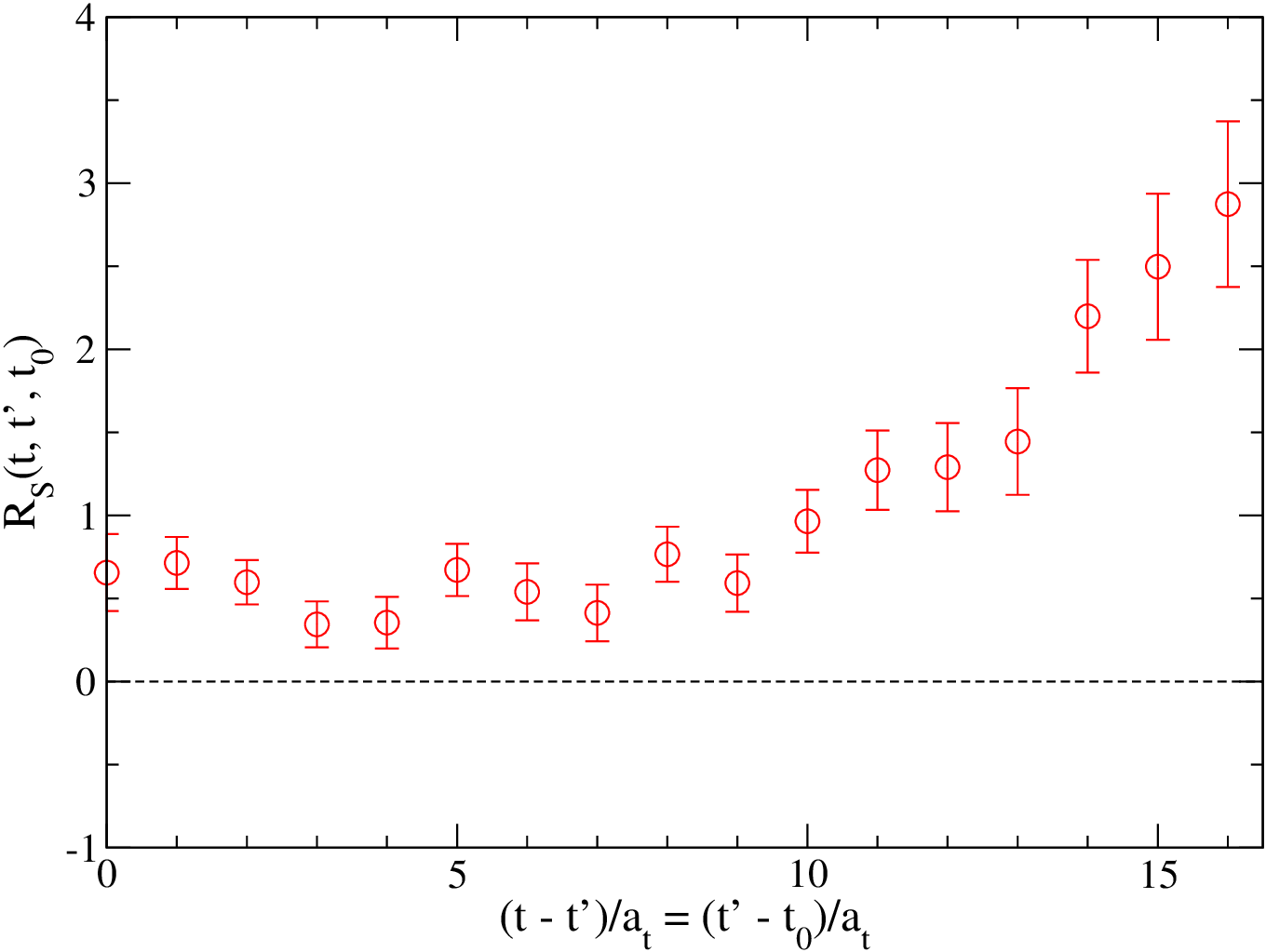}
\caption{\label{fig-scalar}Results for scalar form factor, with $R_S$ as
defined in the text.}
\end{minipage}
\end{figure}

Fig.~\ref{fig-axial} shows our preliminary results for the strange
contribution to the axial form factor, as defined in
Eq.~(\ref{eq-axial}).  Errors have been calculated via jackknife.  We
observe significant correlation at small time separations, but for
$t>5a_t$ the results are consistent with zero.  Our analysis is still
in progress, but it appears that it may only be possible to set a
limit, given the statistics available for this ensemble.  The rapid
disappearance of the signal serves as a strong argument for the
importance of better extended sources for the nucleon.

The prospects for the scalar form factor are more promising, as shown
in Fig.~\ref{fig-scalar}.  Here the signal persists for all time
separations.  The increase at large times is anomalous and possibly
represents a finite size effect, since at $(t-t')=16a_t$ the proton
correlator extends halfway across the lattice.

\begin{figure}
\begin{minipage}[t]{0.48\textwidth}
\includegraphics*[width=\textwidth]{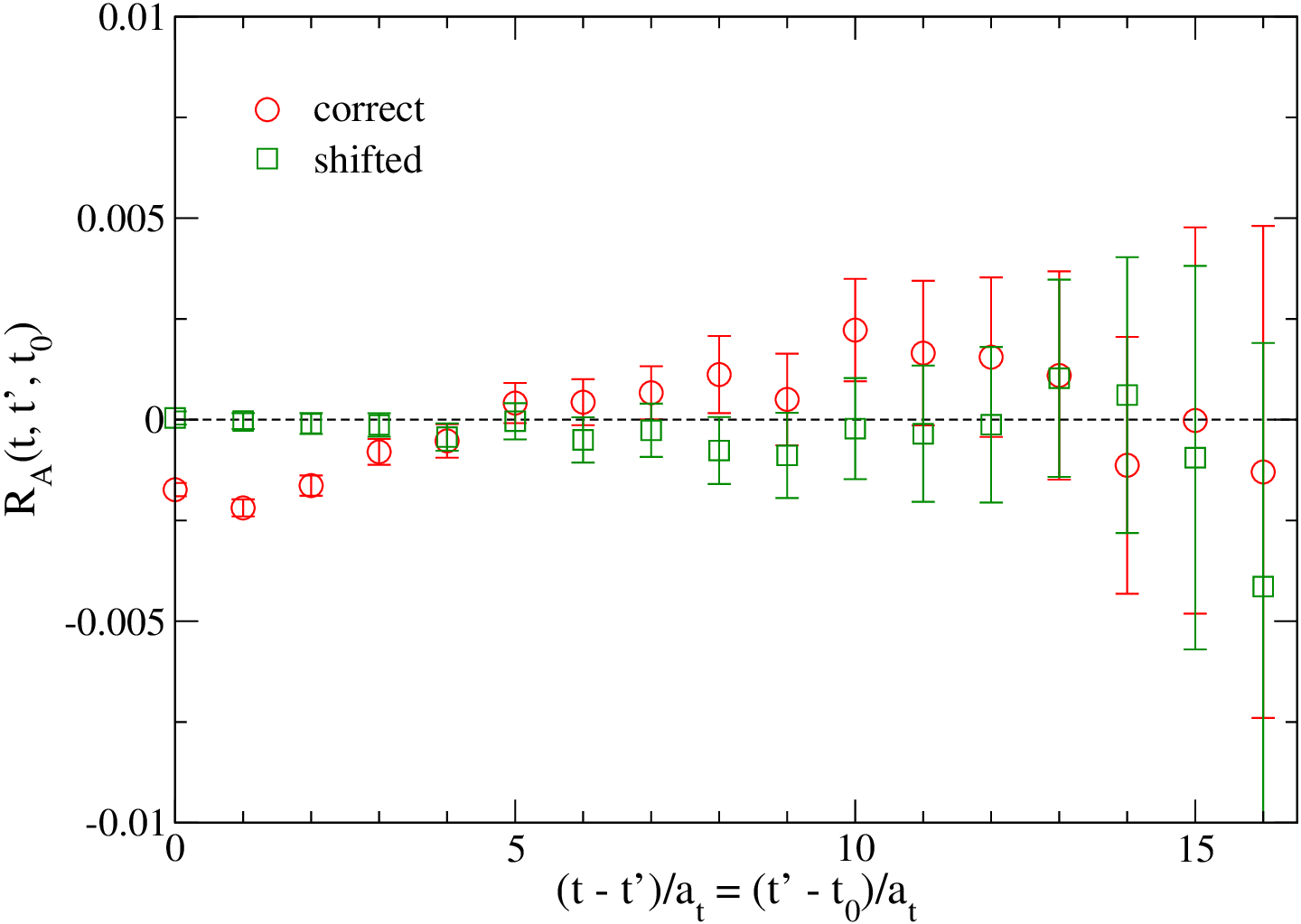}
\caption{\label{fig-axial-shift}Confirmation of correlation in
axial channel.}
\end{minipage}
\hfill
\begin{minipage}[t]{0.48\textwidth}
\includegraphics*[width=0.95\textwidth]{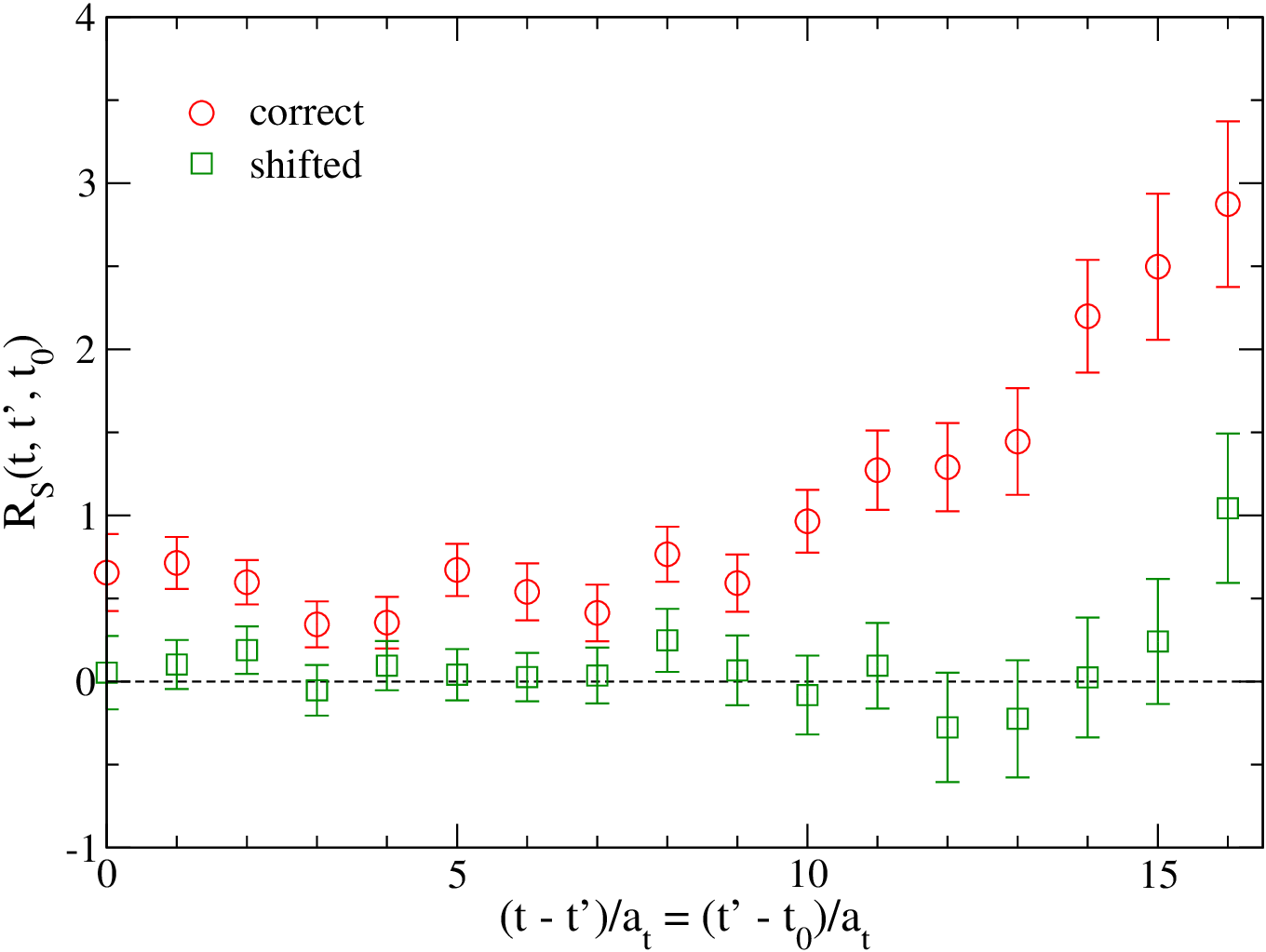}
\caption{\label{fig-scalar-shift}Confirmation of correlation in
scalar channel.}
\end{minipage}
\end{figure}

As shown by these and past results, the evaluation of disconnected
diagrams remains a very challenging problem.  Schematically, we are
calculating a correlation (cf.~Eq.~(\ref{eq-axial}),(\ref{eq-scalar})),
\begin{equation}
\frac{\langle\mathrm{nucleon}\times\mathrm{trace}\rangle
-\langle\mathrm{nucleon}\rangle\langle\mathrm{trace}\rangle}
{\langle\mathrm{nucleon}\rangle}\,,
\end{equation}
that is observed to be small compared to the characteristic
fluctuations in the trace and nucleon correlator.  As a ``sanity
check,'' we confirm that our signal is genuine by shifting
configurations and purposely correlating the nucleon on the $i$th
configuration with the trace on the $(i+s)$th.  As shown in
Figs.~\ref{fig-axial-shift}~and~\ref{fig-scalar-shift}, the signal
vanishes as expected.

\section{Conclusions}

Disconnected diagrams represent a considerable challenge for the
lattice, but they are central to a wide range of physical problems.
Among these is the strange quark contribution to form factors of the
nucleon.  By applying both new and established techniques, we hope to
obtain accurate, unquenched determinations of these quantities.
Preliminary results are encouraging, and should only improve with
increased statistics, better interpolating operators for the nucleon,
and the implementation of multigrid variance reduction.

\acknowledgments

This work was supported in part by US DOE grants DE-FG02-91ER40676 and
DE-FC02-06ER41440 and NSF grants DGE-0221680 and PHY-0427646.  We
thank Boston University and Jefferson Lab for use of their scientific
computing facilities.

\bibliographystyle{JHEPcaps}
\bibliography{lat07_babich}

\end{document}